\documentclass[12pt,a4]{article}
\usepackage{sw20lart}

\input tcilatex
\QQQ{Language}{
American English
}

\begin{document}

\title{\textbf{Chern-Simons as a geometrical set up for three dimensional gauge
theories }}
\author{\textbf{V.E.R. Lemes, C. Linhares de Jesus, } \and \textbf{S.P. Sorella,
L.C.Q.Vilar} \\
UERJ, Universidade do Estado do Rio de Janeiro\\
Departamento de F\'\i sica Te\'orica\\
Instituto de F\'{\i}sica\\
Rua S\~ao Francisco Xavier, 524\\
20550-013, Maracan\~{a}, Rio de Janeiro \vspace{2mm}\\
and\vspace{2mm} \and \textbf{O. S. Ventura} \\
CBPF, Centro Brasileiro\textbf{\ }de Pesquisas F\'{\i}sicas \\
Rua Xavier Sigaud 150, 22290-180 Urca \\
Rio de Janeiro, Brazil\vspace{2mm} \and \textbf{CBPF-NF-078/97}\vspace{2mm}%
\newline \and \textbf{PACS: 11.10.Gh}}
\maketitle

\begin{abstract}
Three dimensional Yang-Mills gauge theories in the presence of the
Chern-Simons action are seen as being generated by the pure topological
Chern-Simons term through nonlinear covariant redefinitions of the gauge
field.

\setcounter{page}{0}\thispagestyle{empty}
\end{abstract}

\vfill\newpage\ \makeatother
\renewcommand{\theequation}{\thesection.\arabic{equation}}

\section{\ Introduction\-}

In a previous letter \cite{lett} it has been observed that the topological
three dimensional massive Yang-Mills gauge theory whose expression is given
by the sum of the Yang-Mills action and of the Chern-Simons term \cite{tmym}

\begin{equation}
\mathcal{S}_{YM}(A)\;+\;\mathcal{S}_{CS}(A)\;,  \label{tmym}
\end{equation}
with

\begin{equation}
\mathcal{S}_{YM}(A)=\frac 1{4m}tr\int d^3xF_{\mu \nu }F^{\mu \nu }\;,
\label{ym}
\end{equation}
and

\begin{equation}
\mathcal{S}_{CS}(A)=\frac 12tr\int d^3x\varepsilon ^{\mu \nu \rho }\left(
A_\mu \partial _\nu A_\rho +\frac 23gA_\mu A_\nu A_\rho \right) \;,
\label{cs}
\end{equation}
can be cast in the form of a pure Chern-Simons action through a nonlinear
redefinition of the gauge connection, namely

\begin{equation}
\mathcal{S}_{YM}(A)\;+\;\mathcal{S}_{CS}(A)=\mathcal{S}_{CS}(\widehat{A})\;,
\label{mf}
\end{equation}
with

\begin{equation}
\widehat{A}_\mu =A_\mu +\sum_{n=1}^\infty \frac 1{m^n}\vartheta _\mu
^n(D,F)\;.  \label{mf1}
\end{equation}
The coefficients $\vartheta _\mu ^n(D,F)$ in eq.$\left( \text{\ref{mf1}}%
\right) $ turn out to be \textit{local }and \textit{covariant}, meaning that
they are built only with the field strength\footnote{%
As usual the gauge field $A_\mu $ is meant to be Lie algebra valued, $A_\mu
=A_\mu ^aT^a$, $T^a$ being the antihermitian generators of a semisimple Lie
group.} $F_{\mu \nu }$

\begin{equation}
F_{\mu \nu }=\partial _\mu A_\nu -\partial _\nu A_\mu +g[A_\mu ,A_\nu ]\;,
\label{fs}
\end{equation}
and the covariant derivative $D_\mu $

\begin{equation}
D_\mu =\partial _\mu +g\left[ A_\mu ,\;\right] \;.  \label{cov-der}
\end{equation}
The two parameters $g,m$ in the above expressions identify the gauge
coupling constant and the so called topological mass \cite{tmym}. According
to the parametrization chosen for the topological massive Yang-Mills action $%
\left( \text{\ref{tmym}}\right) $ we can assign mass dimension 1 to the
gauge field $A_\mu $, so that the parameters $g,m$ are respectively of mass
dimension 0 and 1.

\noindent  For instance, for the first four coefficients of the expansion $%
\left( \text{\ref{mf1}}\right) $ we have \cite{lett}

\begin{eqnarray}
\vartheta _\mu ^1 &=&\frac 14\varepsilon _{\mu \sigma \tau }F^{\sigma \tau
}\;,  \label{coeff} \\
\vartheta _\mu ^2 &=&\frac 18D^\sigma F_{\sigma \mu }\;,  \nonumber \\
\vartheta _\mu ^3 &=&-\frac 1{16}\varepsilon _{\mu \sigma \tau }D^\sigma
D_\rho F^{\rho \tau }\;+\frac g{48}\varepsilon _{\mu \sigma \tau }\left[
F^{\sigma \rho },F_\rho ^{\;\tau }\right] \;,  \nonumber \\
\vartheta _\mu ^4 &=&-\frac 5{128}D^2D^\rho F_{\rho \mu }+\frac 5{128}D^\nu
D_\mu D^\lambda F_{\lambda \nu }\;  \nonumber \\
&&-\frac 7{192}g\left[ D^\rho F_{\rho \tau },F_\mu ^{\;\;\tau }\right]
\;-\frac 1{48}g\left[ D_\nu F_{\mu \lambda },F^{\lambda \nu \;}\right] \;. 
\nonumber
\end{eqnarray}

This work attempts to provide a detailed and self-contained cohomological
analysis of the equation $\left( \text{\ref{mf}}\right) \;$and of the
covariant character of the coefficients $\vartheta _\mu ^n(D,F)$.
Furthermore, the formulas $\left( \text{\ref{mf}}\right) ,\left( \text{\ref
{mf1}}\right) \;$will be generalized to any local\footnote{%
As we shall see in the Sect.5, a rather large class of nonlocal gauge
invariant actions can be contemplated as well.} gauge invariant Yang-Mills
type action\footnote{%
According to the BRST analysis of gauge theories \cite{bdk,dv,bbh}, the name
Yang-Mills type action is employed here to denote a generic integrated local
invariant polynomial with vanishing ghost number built only with the field
strength $F$ and its covariant derivatives. The corresponding actions are of
a very different nature with respect to the Chern-Simons term, which is
known to be BRST invariant only up to a total derivative. It belongs thus to
the so called BRST cohomology modulo $d$, $d$ being the space-time
differential. Although the present considerations are referred to the flat
euclidean space-time, it is worth recalling that while the Chern-Simons term
turns out to be metric independent when defined on curved spaces, the
Yang-Mills type actions couple in a nontrivial and direct way to the
space-time metric \cite{dv}.}: $\int FD^2F,etc.$. These features will enable
us to interpret the topological Chern-Simons term as a gauge invariant
functional acting on a suitably defined space of gauge connections. Any
given local Yang-Mills type action is thus obtained by evaluating the
Chern-Simons functional at a specific point of this space, yielding thus a
pure geometrical set up for the three dimensional gauge theories.

The work is organized as follows. In Sect.2 we review the BRST cohomology of
three dimensional gauge theories in the presence of the Chern-Simons action.
In Sect.3 we establish some general geometrical features of the topological
Chern-Simons term which will account for the covariant character of the
coefficients $\vartheta _\mu ^n\;$and for the eq.$\left( \text{\ref{mf}}%
\right) $. Sect.4 deals with the generalization to a generic Yang-Mills type
action. Although the content of this paper refers mainly to geometrical
aspects, Sect.5 will be devoted to the consequences following from the
relation $\left( \text{\ref{mf}}\right) $ at the quantum level. Further
possible applications will be also outlined.

\section{BRST cohomology of Yang-Mills theory in the presence of the
Chern-Simons term}

\subsection{Generalities}

The BRST cohomology of the Yang-Mills gauge theories has been studied
extensively in the last years. Very general results and theorems have been
established in any space-time dimension \cite{bdk,dv,bbh}, being easily
adapted to the present case. Following the standard procedure \cite{ht,book}%
, the BRST differential $s$ corresponding to the topological massive
Yang-Mills action $\left( \text{\ref{tmym}}\right) $ is given by

\begin{eqnarray}
sA_\mu &=&D_\mu c\;,  \nonumber \\
sc &=&-gc^2\;,  \nonumber \\
sA_\mu ^{*} &=&\frac 12\varepsilon _{\mu \nu \rho }F^{\nu \rho }+\frac
1mD^\nu F_{\mu \nu }-g\left\{ A_\mu ^{*},c\right\} \;,  \nonumber \\
sc^{*} &=&D^\mu A_\mu ^{*}+g\left[ c^{*},c\right] \;,  \label{brst}
\end{eqnarray}
with $c$ being the Faddeev-Popov ghost and $(A_\mu ^{*},c^{*})$ identifying
the two antifields needed in order to implement in cohomology \cite{bbh} the
equations of motion stemming from the action $\left( \text{\ref{tmym}}%
\right) $. The fields and antifields $(A_\mu ,c,A_\mu ^{*},c^{*})$ carry
respectively ghost number $(0,1,-1,-2)$.

In order to provide a cohomological understanding of the equation $\left( 
\text{\ref{mf}}\right) $ we have first to specify the appropriate functional
space for the BRST differential. As suggested by the equation $\left( \text{%
\ref{mf1}}\right) $, the latter will be identified with the space of the
integrated local polynomials in the fields and antifields of arbitrary
dimension. More precisely, the operator $s$ will be allowed to act on the
functional space of the integrated local formal power series in the fields
and antifields. This choice is the most suitable one in view of the
generalization of the eq.$\left( \text{\ref{mf}}\right) $ to higher order
Yang-Mills type actions, $\int FD^2F$, $\int FD^2D^2F$, $etc$., which will
be discussed later on. These terms fit naturally in the space of the local
formal power series in the fields and antifields. Observe also that the
inverse of the topological mass can be interpreted as the expansion
parameter for the formal power series belonging to this functional space, as
in the case of the coefficients $\vartheta _\mu ^n$ of the nonlinear field
redefinition $\left( \text{\ref{mf1}}\right) .\;$

It is rather simple now to convince ourselves that, within the space of the
local formal power series, the presence of the Chern-Simons term in the
initial action $\left( \text{\ref{tmym}}\right) \;$allows us to implement a
recursive procedure which trivializes any BRST invariant term containing
only $F$\ and its covariant derivatives. In order to have a direct and
simple idea of the meaning of this statement it is sufficient to consider
the so called abelian approximation of the BRST transformations $\left( 
\text{\ref{brst}}\right) $, namely

\begin{equation}
s\longrightarrow s_0\;,  \label{ab-brst}
\end{equation}
with

\begin{eqnarray}
s_0A_\mu &=&\partial _\mu c\;,  \nonumber \\
s_0c &=&0\;,  \nonumber \\
s_0A_\mu ^{*} &=&\frac 12\varepsilon _{\mu \nu \rho }F^{0\nu \rho }+\frac
1m\partial ^\nu F_{\mu \nu }^0\;,  \label{abelian} \\
s_0c^{*} &=&\partial ^\mu A_\mu ^{*}\;,  \nonumber  \label{brst-ab}
\end{eqnarray}
and

\begin{equation}
F_{\mu \nu }^0=\partial _\mu A_\nu -\partial _\nu A_\mu \;.  \label{ab-F}
\end{equation}
From eqs.$\left( \text{\ref{abelian}}\right) $ we see that the $s_0-$%
transformations of the fields and antifields correspond to the case in which
all (anti)commutators have been discarded, reducing thus to a set of abelian
transformations. The operator $s_0$ is actually the first term of the
decomposition of the full BRST differential $s$ according to the filtering
operator \cite{dix,book}

\begin{equation}
\mathcal{N}=tr\int d^3x\left( A_\mu \frac \delta {\delta A_\mu }+c\frac
\delta {\delta c}+A_\mu ^{*}\frac \delta {\delta A_\mu ^{*}}+c^{*}\frac
\delta {\delta c^{*}}\right) \;.  \label{filt}
\end{equation}
As it is well known, the relevance of the operator $s_0$ is due to a very
general theorem \cite{dix,book} on the BRST cohomology which states that the
cohomology of the complete BRST differential $s$ is isomorphic to a subspace
of the cohomology of the operator $s_0$. This implies, in particular, that
if the cohomology of $s_0$ is trivial, that of the full operator $s$ will be
empty as well.

Let us now proceed by rewriting the third equation of $\left( \text{\ref
{abelian}}\right) $ in the following form

\begin{equation}
F_{\mu \nu }^0=s_0(\varepsilon _{\mu \nu \rho }A^{*\rho })\;-\frac
1m\varepsilon _{\mu \nu \rho }\partial _\lambda F^{0\rho \lambda }\;,
\label{rec}
\end{equation}
where use has been made of the euclidean normalization

\begin{equation}
\varepsilon ^{\mu \nu \rho }\varepsilon _{\mu \sigma \tau }=\delta _\sigma
^\nu \delta _\tau ^\rho -\delta _\tau ^\nu \delta _\sigma ^\rho \;.
\label{norm}
\end{equation}
From the equation $\left( \text{\ref{rec}}\right) $ we see that we can
replace the field strength $F_{\mu \nu }^0$ by a pure BRST variation with in
addition a term of higher dimension containing a space-time derivative and a
factor $1/m$. The equation $\left( \text{\ref{rec}}\right) $ has the meaning
of a recursive formula since $F_{\mu \nu }^0$ appears on both sides, thereby
allowing us to express $F_{\mu \nu }^0$ as a pure $s_0-$variation, \textit{%
i.e.}

\begin{eqnarray}
F_{\mu \nu }^0 &=&\;s_0(\varepsilon _{\mu \nu \rho }A^{*\rho })\;-\frac
1m\varepsilon _{\mu \nu \rho }\partial _\lambda F^{0\rho \lambda }  \nonumber
\label{f-triv} \\
&=&s_0\left( \varepsilon _{\mu \nu \rho }A^{*\rho }-\frac 1m(\partial _\mu
A_\nu ^{*}-\partial _\nu A_\mu ^{*})\right) +\frac 1{m^2}(\partial _\mu
\partial ^\sigma F_{\nu \sigma }^0-\partial _\nu \partial ^\sigma F_{\mu
\sigma }^0)  \nonumber \\
&=&s_0\left( \varepsilon _{\mu \nu \rho }A^{*\rho }-\frac 1m(\partial _\mu
A_\nu ^{*}-\partial _\nu A_\mu ^{*})+\frac 1{m^2}(\varepsilon _{\mu \nu
\sigma }\partial _\rho -\varepsilon _{\mu \nu \rho }\partial _\sigma
)\partial ^\sigma A^{*\rho }\right) \;  \nonumber \\
&&-\frac 1{m^3}(\varepsilon _{\mu \nu \sigma }\partial _\rho -\varepsilon
_{\mu \nu \rho }\partial _\sigma )\partial ^\sigma \partial _\lambda
F^{0\rho \lambda }  \nonumber \\
&=&s_0\left( \varepsilon _{\mu \nu \rho }A^{*\rho }-\frac 1m(\partial _\mu
A_\nu ^{*}-\partial _\nu A_\mu ^{*})+\frac 1{m^2}(\varepsilon _{\mu \nu
\sigma }\partial _\rho -\varepsilon _{\mu \nu \rho }\partial _\sigma
)\partial ^\sigma A^{*\rho }\right. \;  \nonumber \\
&&\;\;\;\;\;\left. -\frac 1{m^3}(\varepsilon _{\mu \nu \sigma }\partial
_\rho -\varepsilon _{\mu \nu \rho }\partial _\sigma )\varepsilon ^{\rho
\lambda \tau }\partial ^\sigma \partial _\lambda A_\tau ^{*}\right) +O(\frac
1{m^4})  \nonumber \\
&=&.............\;\;.  \label{f-triv}
\end{eqnarray}
It becomes now apparent that this iterative procedure will result in a
formal power series in the expansion parameter $1/m\;$whose coefficients
will contain only the antifield $A_\mu ^{*}\;$and its space-time
derivatives. The above formula expresses the triviality of the field
strength $F_{\mu \nu }^0$. As a consequence, any invariant local term
depending only on $F_{\mu \nu }^0$ and its space-time derivatives can be
written as a pure $s_0-$variation. Of course, the same property holds at the
level of the full BRST operator $s$ with the result that all the invariant
local terms made up with the field strength $F_{\mu \nu }\;$and its
covariant derivatives$\;$can be cast in the form of an exact BRST variation
of a local formal power series. Therefore, from the general results on the
BRST\ cohomology of gauge theories \cite{bdk,dv,bbh}, we can infer that in
the space of the integrated local power series in the fields and antifields
the unique nontrivial element with the quantum numbers of an action can be
identified with the pure topological Chern-Simons term\footnote{%
We recall here that, from the general theorems proven in \cite{bbh}, the
antifields do not contribute to the BRST cohomology in the sector of zero
ghost number for Yang-Mills gauge theories with semisimple group in
arbitrary space-time dimension.} (see also Sects.5,6 of ref.\cite{bh}).

It is worth recalling that, within the BRST algebraic framework, the terms
of the action which are exact turn out to correspond to pure field
redefinitions. Therefore, the formulas $\left( \text{\ref{mf}}\right)
,\left( \text{\ref{mf1}}\right) $ arise as a consequence of the BRST
triviality of the Yang-Mills term. We also underline that the possibility of
rewriting the Yang-Mills action in exact form relies crucially on the
presence of the topological Chern-Simons term in the starting action. As one
can easily understand, this is due to the fact that the field variation of
Chern-Simons yields the (dual) of the field strength $F_{\mu \nu }$, as
expressed by the BRST transformation of the antifield $A_\mu ^{*}$ in eqs.$%
\left( \text{\ref{brst}}\right) $. Without the presence of the term $%
\varepsilon _{\mu \nu \rho }F^{\nu \rho }$ in the right hand side of eqs.$%
\left( \text{\ref{brst}}\right) $ it would be impossible to implement the
previous recursive procedure, as the left hand side of the eq.$\left( \text{%
\ref{rec}}\right) $ would be vanishing. The formula $\left( \text{\ref{rec}}%
\right) $ would become thus useless. This means that if the Chern-Simons
term is not included in the initial action, there is no way of
(re)expressing the Yang-Mills action in the form of an exact variation of a
local formal power series. However, as soon as the topological Chern-Simons
is turned on, we can immediately reabsorb the Yang-Mills term through a
nonlinear field redefinition.

\subsection{Complete ladder structure}

The previous cohomological considerations can be understood in a simple way
by noticing that the transformations $\left( \text{\ref{abelian}}\right) $
can be cast in a form which is typical of the topological theories of the
Schwartz type \cite{bbrt}, as for instance pure Chern-Simons. In fact, using
as new variables the redefined antifields

\begin{eqnarray}
\widetilde{A}_{\mu \nu }^{*} &=&A_{\mu \nu }^{*}-\frac 1m\varepsilon _{\mu
\nu \rho }\partial _\sigma A^{*\rho \sigma }-\frac 1{m^2}\partial ^2A_{\mu
\nu }^{*}+\frac 1{m^3}\varepsilon _{\mu \nu \rho }\partial ^2\partial
_\sigma A^{*\rho \sigma }  \label{red-ant} \\
&&+\frac 1{m^4}\partial ^2\partial ^2A_{\mu \nu }^{*}+O(1/m^5)\;,  \nonumber
\\
\widetilde{c}^{*} &=&c^{*}-\frac 1{m^2}\partial ^2c^{*}+\frac 1{m^4}\partial
^2\partial ^2c^{*}+O(1/m^5)\;,  \nonumber
\end{eqnarray}
with

\begin{equation}
A_{\mu \nu }^{*}=\varepsilon _{\mu \nu \rho }A^{*\rho }\;,  \label{d-ant}
\end{equation}
one easily gets, up to the order $1/m^5$,

\begin{eqnarray}
s_0A_\mu &=&\partial _\mu c\;,  \label{ab-c-l} \\
s_0c &=&0\;,  \nonumber \\
s_0\widetilde{A}_{\mu \nu }^{*} &=&F_{\mu \nu }^0\;,  \nonumber \\
s_0\widetilde{c}^{*} &=&\frac 12\varepsilon ^{\mu \nu \rho }\partial _\mu 
\widetilde{A}_{\nu \rho }^{*}\;.  \nonumber  \label{brst-ab}
\end{eqnarray}
This structure, called complete ladder structure \cite{book}, implies that
all fields but the undifferentiated ghost $c$ can be grouped in BRST
doublets, meaning that the cohomology of $s_0$ in the space of the formal
power series is spanned by polynomials in the undifferentiated ghost $c.$ As
it is well known this result, combined with the requirement of the rigid
gauge invariance \cite{bdk,bbh}, allows to identify the cohomology classes
of the full BRST differential with the invariant polynomials in the
undifferentiated Faddeev-Popov ghost $c$ built with monomials of the kind $%
trc^{2n+1}$, $n\ge 1$. It follows then that the cohomology of $s$ modulo $d$
in the sector of the local power series with the same quantum numbers of a
Lagrangian has a unique nontrivial element, corresponding (via descent
equations \cite{book}) to the ghost monomial $trc^3$. The resulting action
is the Chern-Simons term.

Having justified the equations $\left( \text{\ref{mf}}\right) ,\left( \text{%
\ref{mf1}}\right) $, let us now turn to the covariant character of the
coefficients $\vartheta _\mu ^n$ in eqs.$\left( \text{\ref{mf1}}\right) $.
This will be the task of the next Section.

\section{Some useful properties of the pure Chern-Simons action}

In order to account for the covariant character of the coefficients $%
\vartheta _\mu ^n$ in the eq.$\left( \text{\ref{mf1}}\right) $ we recall
first some simple properties of the Chern-Simons term. Let $A_\mu \;$be a
given gauge connection and let $\mathcal{S}_{CS}(A)$ be the corresponding
BRST invariant Chern-Simons action, as given by the expression $\left( \text{%
\ref{cs}}\right) $. Let us now vary the gauge field $A_\mu $ by an arbitrary
amount $\delta A_\mu $ and let us try to establish the transformation law
for $\delta A_\mu $ in order that the new Chern-Simons functional $\mathcal{S%
}_{CS}(\widehat{A})\;$evaluated at $\widehat{A}_\mu =A_\mu +\delta A_\mu $, 
\textit{i.e.}

\begin{equation}
\mathcal{S}_{CS}(\widehat{A})=\mathcal{S}_{CS}(A)+tr\int d^3x\varepsilon
^{\mu \nu \rho }\left( \frac 12\delta A_\mu F_{\nu \rho }+\frac 12\delta
A_\mu D_\nu \delta A_\rho +\frac g3\delta A_\mu \delta A_\nu \delta A_\rho
\right) \;,  \label{ncs}
\end{equation}
is still BRST invariant. Notice also that the variation $\delta A_\mu $ is
not treated as an \textit{infinitesimal }quantity, the formula $\left( \text{%
\ref{ncs}}\right) $ being indeed exact.

\noindent Requiring then that

\begin{equation}
s\mathcal{S}_{CS}(\widehat{A})=0\;,  \label{inv-ncs}
\end{equation}
and recalling that

\begin{equation}
s\mathcal{S}_{CS}(A)=0\;,  \label{inv-cs}
\end{equation}
we easily obtain

\begin{eqnarray}
0 &=&tr\int d^3x\varepsilon ^{\mu \nu \rho }\left( \left( s\delta A_\mu
-g\left[ \delta A_\mu ,c\right] \right) F_{\nu \rho }+2g(s\delta A_\mu
)\delta A_\nu \delta A_\rho \right.  \nonumber  \label{con-ncs} \\
&&\;\;\;\;\;\;\;\;\;\;\;\;+\;\left( (s\delta A_\mu )D_\nu \delta A_\rho
+\delta A_\mu s(D_\nu \delta A_\rho )\right) \left. {}\right) \;.
\label{con-ncs}
\end{eqnarray}
The condition $\left( \text{\ref{con-ncs}}\right) \;$implies that

\begin{equation}
s\delta A_\mu =g\left[ \delta A_\mu ,c\right] \;,  \label{cov-con}
\end{equation}
meaning thus that $\delta A_\mu $ transforms covariantly. From the eq.$%
\left( \text{\ref{cov-con}}\right) $ it follows that the modified field $%
\widehat{A}_\mu =A_\mu +\delta A_\mu $ is a \textit{connection},

\begin{equation}
s\widehat{A}_\mu =\partial _\mu c+g\left[ \widehat{A}_\mu ,c\right] \;,
\label{conn}
\end{equation}
as it should be.

\noindent We see therefore that if we vary the gauge field $A_\mu \;$by an
arbitrary amount $\delta A_\mu \;$which transforms covariantly under BRST,
the resulting Chern-Simons term $\mathcal{S}_{CS}(A_\mu +\delta A_\mu )\;$%
will remain gauge invariant. Of course, the covariant character persists
also in the case in which $\delta A_\mu $ is meant to be a local formal
power series in the expansion parameter $1/m$, \textit{i.e.}

\begin{equation}
\delta A_\mu =\sum_{n=1}^\infty \frac 1{m^n}\vartheta _\mu ^n\;.
\label{dafp}
\end{equation}
From the eq.$\left( \text{\ref{cov-con}}\right) \;$we have

\begin{equation}
s\vartheta _\mu ^n=g\left[ \vartheta _\mu ^n,c\right] \;,  \label{th-cov}
\end{equation}
owing to the fact that coefficients $\vartheta _\mu ^n\;$with different
values of $n$ have to be considered independent, being of different mass
dimensions.

\noindent Moreover, from the eq.$\left( \text{\ref{dafp}}\right) $ we
obviously get

\begin{equation}
\mathcal{S}_{CS}(A_\mu +\delta A_\mu )=\mathcal{S}_{CS}(A)+\sum_{n=1}^\infty
\frac 1{m^n}\mathcal{S}^n\;,  \label{ncs-exp}
\end{equation}
$\mathcal{S}^n\;$being integrated local formal power series corresponding to
the expansion of $\mathcal{S}_{CS}(A_\mu +\delta A_\mu )$ in powers of the
inverse of the topological mass $m$, according to eq.$\left( \text{\ref{dafp}%
}\right) $.

\noindent Furthermore, from the BRST invariance of $\mathcal{S}_{CS}(A_\mu
+\delta A_\mu )$ and of $\mathcal{S}_{CS}(A)$, we have

\begin{equation}
s\mathcal{S}^n=0\;,  \label{inv-csnn}
\end{equation}
implying that the coefficients $\mathcal{S}^n$ in the eq.$\left( \text{\ref
{ncs-exp}}\right) $ are BRST invariant. Recalling now that the Chern-Simons
term is the unique nontrivial action in the space of the local formal power
series, it follows that the $\mathcal{S}^n$'s in eq.$\left( \text{\ref
{ncs-exp}}\right) $ have to be necessarely BRST exact, namely

\begin{equation}
\mathcal{S}^n=s\widehat{\mathcal{S}\,}^n\,,  \label{triv-csnn}
\end{equation}
for some local integrated formal power series $\widehat{\mathcal{S}\,}^n\;$%
with negative ghost number. We are now ready to give a cohomological proof
of the equation $\left( \text{\ref{mf}}\right) $.

\noindent In fact, the following Lemma holds:

\begin{lemma}
Among the class of the BRST invariant Chern-Simons functionals $\mathcal{S}%
_{CS}(\widehat{A})$%
\begin{equation}
\mathcal{S}_{CS}(\widehat{A})=\frac 12tr\int d^3x\varepsilon ^{\mu \nu \rho
}\left( \widehat{A}_\mu \partial _\nu \widehat{A}_\rho +\frac 23g\widehat{A}%
_\mu \widehat{A}_\nu \widehat{A}_\rho \right) \;,  \label{cs-funct}
\end{equation}
$\widehat{A}_\mu $ being a connection of the type 
\begin{equation}
\;\widehat{A}_\mu =A_\mu \;+\sum_{n=1}^\infty \frac 1{m^n}\vartheta _\mu
^n\;,  \label{cs-funct-ex}
\end{equation}

\noindent it is always possible to find a set of covariant\footnote{%
We recall that the covariant character of the $\vartheta _\mu ^n$'s follows
from the requirement of gauge invariance of ${\mathcal{S}}_{CS}(\hat{A})$.}
coefficients $\vartheta _\mu ^n\;$such that 
\begin{equation}
\mathcal{S}_{CS}(\widehat{A})=\mathcal{S}_{CS}(A)+\frac 1{4m}tr\int
d^3xF_{\mu \nu }F^{\mu \nu }\;.  \label{triv-ym}
\end{equation}

\textbf{Proof.}

In order to prove the Lemma we proceed by assuming the converse, as it is
usual in this kind of problem. Let us suppose then that the eq.$\left( \text{%
\ref{triv-ym}}\right) \;$does not hold, i.e. that 
\begin{equation}
\mathcal{S}_{CS}(\widehat{A})\neq \mathcal{S}_{CS}(A)+\frac 1{4m}tr\int
d^3xF_{\mu \nu }F^{\mu \nu }\;.  \label{nt}
\end{equation}
Therefore, recalling from the eq.$\left( \text{\ref{triv-csnn}}\right) \;$%
that 
\begin{equation}
\mathcal{S}_{CS}(\widehat{A})=\mathcal{S}_{CS}(A)+s\left( \sum_{n=1}^\infty
\frac 1{m^n}\widehat{\mathcal{S}\,}^n\right) \;,  \label{cs-csnn}
\end{equation}
we should have 
\begin{equation}
\frac 1{4m}tr\int d^3xF_{\mu \nu }F^{\mu \nu }\neq s\left( \sum_{n=1}^\infty
\frac 1{m^n}\widehat{\mathcal{S}\,}^n\right) \;,  \label{cont}
\end{equation}
which, of course, is in contrast with the results of the previous Section
which allow us in fact to express the Yang-Mills action as a pure BRST
variation of a local formal power series, thereby concluding the proof of
the Lemma.
\end{lemma}

\vspace{5mm}

The above result provides a simple cohomological understanding of the
equations $\left( \text{\ref{mf}}\right) ,\left( \text{\ref{mf1}}\right) $.
It can be easily extended to cover the case in which the initial action $%
\left( \text{\ref{tmym}}\right) $ is supplemented with generalized terms of
the Yang-Mills type. We recall in fact that among the BRST invariant actions
built with $F_{\mu \nu }$ and its covariant derivatives, the Yang-Mills
Lagrangian $trF_{\mu \nu }F^{\mu \nu }$ is the term with the lowest mass
dimension. Any other term of this kind will contain a higher number of $%
F_{\mu \nu }\;$or $D_\mu $, increasing thus its mass dimension. As a
consequence, the abelian transformations $\left( \text{\ref{abelian}}\right) 
$ will get modified by terms of higher order in $F_{\mu \nu }^0\;$and its
space-time derivatives. Therefore, provided the Chern-Simons term is
included in the starting action, it will be always possible to generalize
the formula $\left( \text{\ref{rec}}\right) $, thereby expressing the field
strength $F_{\mu \nu }^0$ as a BRST exact variation of a local formal power
series. Everything will work as before, with the only difference that the
coefficients $\vartheta _\mu ^n$ of the nonlinear redefinition $\left( \text{%
\ref{mf1}}\right) $ have now to be suitably modified. However they will
remain covariant, as it will be illustrated in the following examples.

\section{Examples}

In order to have a better understanding of the previous results it is useful
to work out the expressions of the coefficients $\vartheta _\mu ^n\;$in the
case in which we add to the initial action $\left( \text{\ref{tmym}}\right) $
generalized terms of the Yang-Mills type. We shall study in particular the
following terms

\begin{equation}
\mathcal{S}_\lambda (A)=\frac \lambda {2m^2}tr\int d^3x\varepsilon ^{\mu \nu
\rho }F_{\mu \sigma }D_\nu F_\rho ^{\;\sigma }\;,  \label{ex1}
\end{equation}
and

\begin{equation}
\mathcal{S}_\tau (A)=\frac \tau {4m^3}tr\int d^3xF^{\mu \nu }D^2F_{\mu \nu
}\;,  \label{ex2}
\end{equation}
$\lambda ,\tau \;$being two dimensionless arbitrary parameters. The terms $%
\left( \text{\ref{ex1}}\right) ,\left( \text{\ref{ex2}}\right) $\ have been
considered in fact by \cite{g1,as}$\;$as higher derivatives regularizing
actions for the topological massive Yang-Mills.

\noindent Let us choose then as initial action the expression

\begin{equation}
\mathcal{S}_{YM}(A)\;+\;\mathcal{S}_{CS}(A)+\mathcal{S}_\lambda (A)\;.
\label{act1}
\end{equation}
In this case, for the first coefficients $\vartheta _\mu ^n\;$of the
expansion $\left( \text{\ref{cs-funct-ex}}\right) $ we get

\begin{eqnarray}
\vartheta _\mu ^1 &=&\frac 14\varepsilon _{\mu \sigma \tau }F^{\sigma \tau
}\;,  \nonumber  \label{coeff} \\
\vartheta _\mu ^2 &=&\frac{(1-4\lambda )}8D^\sigma F_{\sigma \mu }\;, 
\nonumber \\
\vartheta _\mu ^3 &=&-\frac{(1-4\lambda )}{16}\varepsilon _{\mu \sigma \tau
}D^\sigma D_\rho F^{\rho \tau }\;+\frac g{48}\varepsilon _{\mu \sigma \tau
}\left[ F^{\sigma \rho },F_\rho ^{\;\tau }\right] \;.  \label{coeff1}
\end{eqnarray}
Analogously, in the case in which the starting point is

\begin{equation}
\mathcal{S}_{YM}(A)\;+\;\mathcal{S}_{CS}(A)+\mathcal{S}_\tau (A)\;,
\label{act2}
\end{equation}
we obtain

\begin{eqnarray}
\vartheta _\mu ^1 &=&\frac 14\varepsilon _{\mu \sigma \tau }F^{\sigma \tau
}\;,  \nonumber  \label{coeff} \\
\vartheta _\mu ^2 &=&\frac 18D^\sigma F_{\sigma \mu }\;,  \nonumber \\
\vartheta _\mu ^3 &=&-\frac 1{16}\varepsilon _{\mu \sigma \tau }D^\sigma
D_\rho F^{\rho \tau }\;+\frac g{48}\varepsilon _{\mu \sigma \tau }\left[
F^{\sigma \rho },F_\rho ^{\;\tau }\right] \;+\frac \tau 4\varepsilon _{\mu
\nu \rho }D^2F^{\nu \rho }\;.  \label{coeff2}
\end{eqnarray}
Of course, provided the Chern-Simons term is included in the starting
action, any other combination of $\int trF^2,\;\int trFD^2F,\;\int
tr\varepsilon ^{\mu \nu \rho }F_{\mu \sigma }D_\nu F_\rho ^{\;\sigma },\;$%
will lead to similar results. Notice, finally, that the expression for the
coefficient $\vartheta _\mu ^1\;$is independent from the parameters $\lambda
,\tau $, owing to the fact that the two terms $\mathcal{S}_\lambda (A)\;$and 
$\mathcal{S}_\tau (A)$ in eqs.$\left( \text{\ref{ex1}}\right) ,\left( \text{%
\ref{ex2}}\right) $ are respectively of the order $1/m^2$ and $1/m^3$.

\section{Conclusion}

The results established in the previous sections lead us to organize the
concluding remarks in two separate classes. To the first class belong the
pure geometrical considerations. In the second one we discuss the field
theory aspects. Here we will attempt to make contact with the known
perturbative results on topological massive Yang-Mills \cite
{tmym,pr,gmrr,as,gl,kp,rec}. We shall also try to provide a meaningful
support to a question which arises almost naturally and which could be of
great interest in order to improve our present understanding of the
effective $1PI$ quantum actions of three dimensional gauge theories.

\subsection{The geometrical set up}

A very simple and attractive geometrical set up emerges from the
considerations of this work. The Chern-Simons term can be interpreted in
fact as a gauge invariant functional defined on the space of all possible
gauge connections of the kind $\left( \text{\ref{cs-funct-ex}}\right) $. Any
given local Yang-Mills type action is then reproduced by evaluating the
Chern-Simons functional at a specific point of this space, which amounts to
a suitable choice of the gauge connection or, equivalently, of the covariant
coefficients $\vartheta _\mu ^n$. In this sense the Chern-Simons term may be
considered as a topological generator for three dimensional Yang-Mills gauge
theories.

Moreover, this geometrical interpretation gives us a direct perception of
how \textit{rigid} can be a topological object. Of course, rigidity has here
the meaning of the classical equivalence, up to local nonlinear field
redefinitions, among the Yang-Mills type actions in the presence of
Chern-Simons and the pure Chern-Simons term. It is worth noticing that the
aforementioned rigidity of Chern-Simons has been already underlined by \cite
{bh} in the framework of the consistent deformations of the master equation.

\subsection{Field theory aspects}

\subsubsection{Perturbative topological massive Yang-Mills and ultraviolet
finiteness}

In spite of being only power counting superrenormalizable, topological
massive Yang-Mills $\left( \text{\ref{tmym}}\right) $ is ultraviolet finite
to all orders of perturbation theory. This property has been first observed
at one loop level \cite{tmym,pr} and later on has been extended to all
orders by combining two loops computations with finiteness by power counting
at higher loops \cite{gmrr}. The ultraviolet finiteness has been proven in
the Landau gauge; this gauge being always assumed in what follows.

In order to make use of the results established in Sects.3,4, let us first
analyse the meaning of the formulas $\left( \text{\ref{triv-ym}}\right)
,\left( \text{\ref{cs-csnn}}\right) $ from a field theory point of view. In
particular, the eq.$\left( \text{\ref{cs-csnn}}\right) $ implies that the
Yang-Mills action can be written as a pure BRST variation of a power series
which contains terms of arbitrary dimension, according to the nonlinearity
of the field redefinition $\left( \text{\ref{mf1}}\right) $. This could seem
to be in disagreement with the standard power counting, since the BRST
differential is required to act on the space of local terms of arbitrary
dimension. Nevertheless, we can give a meaning to eq.$\left( \text{\ref
{cs-csnn}}\right) $ by adopting a more general point of view and take as the
starting action the formal power series

\begin{equation}
\mathcal{S}(A)=\mathcal{S}_{CS}(A)+\frac 1{4m}tr\int d^3xF_{\mu \nu }F^{\mu
\nu }+\sum_{j=2}^\infty \frac 1{m^j}\left( \sum_{k=1}^{d_k}\alpha _k^j%
\mathcal{S}_j^k(A)\right) \;,  \label{m-g}
\end{equation}
where $\alpha _k^j$ are arbitrary coefficients and $\mathcal{S}_j^k$ are all
possible higher dimensional Yang-Mills type actions built with the field
strength $F$ and the covariant derivative $D$. The index $k$ in the double
sum $\left( \text{\ref{m-g}}\right) $ is needed in order to account for the
degeneracy $(d_k)$ of different Yang-Mills actions with the same dimension.
Notice also that, according to the results of Sects.3,4, the action ${%
\mathcal{S}}(A)$ in eq.$\left( \text{\ref{m-g}}\right) $ can be cast in the
form of a pure Chern-Simons term, \textit{i.e.} ${\mathcal{S}}(A)={\mathcal{S%
}}_{CS}(\widehat{A})$, with a suitable choice of the gauge connection $%
\widehat{A}$.

\noindent For the fully quantized action in the Landau gauge we have
therefore

\begin{equation}
\Sigma =\mathcal{S}+tr\int d^3x\left( b\partial A+\partial ^\mu \overline{c}%
D_\mu c+A_\mu ^{*}D^\mu c-gc^{*}c^2\right) \;,  \label{ca-l}
\end{equation}
where $b,\overline{c}$\ are the lagrangian multiplier and the antighost. The
reason for this choice is that, being now the starting action $\Sigma $ a
formal power series, the BRST differential turns out to be naturally defined
on the space of the local formal power series.

It is worth underlining that this point of view closely follows the recent
new perspectives on the renormalization of gauge theories outlined in \cite
{gw}. Owing to the general results on the cohomology of gauge theories \cite
{bdk,bbh}, the action $\left( \text{\ref{ca-l}}\right) $ is indeed
renormalizable\footnote{%
According to the analysis of \cite{gw}, the model ($\ref{ca-l})$ can be seen
as a renormalizable theory fulfilling the so called structural constraint of
the type A.} in the sense that all divergences can be cancelled by the
infinite terms of $\Sigma $, which spans all possible local BRST invariant
Yang-Mills terms.

Although the action $\Sigma $ justifies the use of the space of the formal
power series for the BRST differential, we have always to face the problem
of the infinite number of parameters present in the expression $\left( \text{%
\ref{m-g}}\right) $. However, it is easily seen that all the coefficients $%
(m,\alpha _k^j)$ correspond to BRST trivial parameters \cite{book}. Indeed,
recalling that from the results of the previous sections every Yang-Mills
type term can be written as the BRST variation of a formal power series, we
immediately infer that

\begin{eqnarray}
\frac{\partial \Sigma }{\partial m} &=&\;BRST-\mathrm{variation}\;,\;
\label{brst-v} \\
\frac{\partial \Sigma }{\partial \alpha _k^j} &=&\;BRST-\mathrm{variation\;},
\nonumber
\end{eqnarray}
the left hand side of eq.$\left( \text{\ref{brst-v}}\right) $ being
understood as a formal power series. \textit{The above equations mean thus
that the dependence of the action }$\mathcal{S}$\textit{\ from the
parameters }$m,\alpha _k^j$\textit{\ can be controlled through a nonlinear
redefinition of the gauge field, as showed in Sects.3,4.}

We see therefore that, in spite of the presence of an infinite number of
parameters, the eq.$\left( \text{\ref{brst-v}}\right) $ implies that $\Sigma
\;$possesses a unique nontrivial parameter\footnote{%
This situation looks rather similar to that of other kinds of well known
models, as for instance the two dimensional nonlinear sigma model \cite
{sigma} and the superspace four dimensional $N=1$ super Yang-Mills \cite{n1}
which, in spite of the presence of an infinite number of parameters, turn
out to be characterized only by a finite set of BRST nontrivial couplings.
It should be noticed also that the requirement that only a finite number of
parameters are BRST nontrivial is a condition stronger than those assumed in 
\cite{gw}. This requirement, combined with the fulfillment of the structural
constraints of \cite{gw}, could give a more precise meaning to theories
which are apparently powercounting nonrenormalizable.} $g$, corresponding
indeed to the Chern-Simons action,

\begin{equation}
g\frac{\partial \Sigma }{\partial g}=\mathcal{S}_{CS}(A)+\left( BRST-\mathrm{%
variation}\right) \mathrm{\;}.  \label{g-nt}
\end{equation}
In addition, due to the topological character of Chern-Simons, it is
apparent that an equation similar to $\left( \text{\ref{brst-v}}\right) $
holds for the classical BRST invariant symmetric energy momentum tensor%
\footnote{%
As it is well known, a symmetric classical BRST invariant energy-momentum
tensor $T_{\mu \nu }$ can be obtained by the standard procedure of coupling
the action $\mathcal{S}$ to gravity and set the metric to the flat one after
taking the derivative of $\mathcal{S}$, \textit{i.e.}
\par
\[
T_{\mu \nu }=\frac 1{\sqrt{\eta }}\left. \frac{\delta \mathcal{S}}{\delta
\eta ^{\mu \nu }}\right| _{\eta \rightarrow \;flat}\;. 
\]
The Chern-Simons term does not contribute to $T_{\mu \nu }$ due to the fact
that it does not couple to the metric.} $T_{\mu \nu }\;$computed from $%
\mathcal{S},$

\begin{equation}
T_{\mu \nu }=BRST-\mathrm{variation\;.}  \label{e-m}
\end{equation}
Furthermore, making use of the extended BRST technique \cite{co} (see App.A
for the details) and bearing in mind that in three dimensions there are no
gauge anomalies, it follows that the equations $\left( \text{\ref{brst-v}}%
\right) ,\left( \text{\ref{e-m}}\right) $ are easily extended at the quantum
level, namely 
\begin{eqnarray}
\frac{\partial \Gamma }{\partial m} &=&\;BRST-\mathrm{variation}\;,
\label{q-e-m} \\
\frac{\partial \Gamma }{\partial \alpha _k^j} &=&\;BRST-\mathrm{variation\;},
\nonumber \\
\left[ T_{\;\mu }^\mu \cdot \Gamma \right] &=&BRST-\mathrm{%
variation\;+\;total}\;\mathrm{derivative\;},  \nonumber
\end{eqnarray}
where, as usual, the terms in the left hand side have to be understood as
BRST\ exact quantum insertions of formal power series (see App.A) and $%
T_{\;\mu }^\mu \;$stands for the trace of $T_{\mu \nu }.$

This implies, in particular, that the dependence of the quantum theory from
the renormalization point $\mu $ can be controlled by the introduction of
suitable BRST exact terms. These terms, being formal power series,
correspond to possible nonlinear redefinitions of the fields.

The equations $\left( \text{\ref{q-e-m}}\right) $, implicitly contained in
the analysis of ref.\cite{lett}, represent our algebraic understanding of
the ultraviolet finiteness properties and of the role of the nonlinear field
redefinitions at the quantum level.

Recently, an independent and different proof of the ultraviolet finiteness
of topological massive Yang-Mills including the vanishing of the field
anomalous dimensions has been achieved by \cite{rec}.

\subsubsection{An open question: the fully quantum equivalence}

Perhaps the most intriguing question which arises naturally in the present
context is whether the classical equivalence between the massive topological
Yang-Mills and the pure Chern-Simons, as stated by eq.$\left( \ref{mf}%
\right) $, can be extended at the level of the full $1PI$ effective action $%
\Gamma (A).$

Of course, we are unable to give a satisfactory and definitive answer to
this question. Only a few terms of $\Gamma (A)$ have been computed till now; 
$\Gamma (A)$ being an infinite series in the loop parameter expansion $\hbar 
$.

What we shall do here is to show that, in spite of their \textit{nonlocal }%
character\textit{, }a rather large number of terms contributing to $\Gamma
(A)$ can be in fact reabsorbed in the Chern-Simons through \textit{nonlocal }%
but \textit{covariant} nonlinear field redefinitions.

In what follows we shall strictly refer to the flat euclidean space-time $%
\mathcal{R}^3$ endowed with the constant flat metric $\eta _{\mu \nu }=%
\mathrm{diag.}(+,+,+).$

We begin with a more precise definition of the $1PI$ quantum action. The
functional $\Gamma (A)$ is meant to be the $1PI$\ effective action computed
from topological massive Yang-Mills upon quantization in the Landau gauge
and after setting to zero the classical fields corresponding to the
lagrangian multiplier, ghosts and antifields. Therefore $\Gamma (A)$ depends
only on the classical field $A_\mu \;$defined through the Legendre
transformation of the generator $\mathcal{Z}^c(J)$ of the connected Green's
functions, \textit{i.e.}

\begin{equation}
\Gamma (A)=\sum_{n=2}^\infty \int d^3x_1....d^3x_nA(x_1)....A(x_n)\Gamma
^n(x_{1,....,}x_n)\;,  \label{1pi}
\end{equation}
where $\Gamma ^n(x_{1,....,}x_n)$ is the $n$-point\footnote{%
We have not specified the group indices in eq.$\left( \ref{1pi}\right) $
since they are not relevant for the forthcoming considerations.} $1PI$ Green
function.

Let us recall here the main basic facts about pure Chern-Simons and
topological massive Yang-Mills at the quantum level which will be useful in
the following:

\begin{itemize}
\item  $\Gamma (A)$ is gauge invariant and contains both local and \textit{%
nonlocal }contributions at each order of perturbation theory.

\item  For a pure Chern-Simons theory the general covariance is unbroken at
the quantum level \cite{en,ag,al,noi,lu}, implying that the dependence on
the (flat) space-time metric is nonphysical. Let us also remind that,
although being only powercounting renormalizable, Chern-Simons is
ultraviolet finite \cite{en,ag,al,noi}.

\item  As proven by \cite{gmrr}, pure Chern-Simons is recovered as the
infinite mass limit $m\rightarrow \infty $ of topological massive Yang-Mills.

\item  In spite of the presence of the antisymmetric tensor $\varepsilon
_{\mu \nu \rho }$, explicit regularizations preserving the BRST invariance
have been constructed for topological massive Yang-Mills \cite{gmrr,as,gl}.
Although not needed for an algebraic analysis, the existence of
regularizations preserving the BRST symmetry is rather important in order to
carry out computations. This means that the BRST invariance can be
maintained manifestly at each step, implying that $\Gamma (A)$ can be
constructed without leaving the space of the gauge invariant terms.
\end{itemize}

\noindent Let us now focus on the structure of the $1PI$ effective action $%
\Gamma (A)$. Needless to say, $\Gamma (A)\;$is expected to be a quite
complicated object. Moreover, due to the gauge invariance, $\Gamma (A)\;$%
will certainly contain a large number of terms built with the field strength 
$F$ and its covariant derivatives, intertwined in very complicated nonlocal
but gauge invariant combinations \cite{pr}.

However, it is very simple to see that if we start with a gauge invariant
nonlocal action built with $F$ and its covariant derivatives in the presence
of Chern-Simons, the recursive formula $\left( \text{\ref{rec}}\right) $ can
be suitably adapted to the nonlocal case, with the result that the nonlocal
term can be reabsorbed in the pure Chern-Simons through a \textit{nonlocal}
field redefinition. Remarkably in half and in spite of the nonlocality, the
redefined gauge field will still transform as a connection, due to the fact
that the coefficients entering the nonlocal redefinition turn out to be
covariant, as it will be shown in the next example. Thus, a great amount of
nonlocal quantum effects coming from nonlocal terms built with $F$ can be
taken into account by interpreting the Chern-Simons as a gauge invariant
functional defined on the space of the gauge connections of the type $%
\widehat{A}=A+\delta A$, where $\delta A$ contains now both local and
nonlocal covariant terms, $\delta A=\delta A^{loc}+\delta A^{nloc}.\;$

In order to have a better feeling of how things go in the nonlocal case, let
us compute the first coefficients $\vartheta _\mu ^1,\vartheta _\mu ^2$ for
the nonlocal action

\begin{equation}
\mathcal{S}_{YM}^{nloc}(A)=\frac 1{4m}\int d^3xd^3yF^2(x)\left| x-y\right|
F^2(y)\;,  \label{nl-ym}
\end{equation}
with

\begin{equation}
F^2(x)=trF^{\mu \nu }(x)F_{\mu \nu }(x)\;.  \label{tr}
\end{equation}
The expression $\left( \text{\ref{nl-ym}}\right) $ is one of the simplest
nonlocal invariant terms depending on $F$ which is expected to appear in the
loop expansion of $\Gamma (A)$. The coefficients $\vartheta _\mu
^1,\vartheta _\mu ^2$ are easily found to be

\begin{eqnarray}
\vartheta _\mu ^1 &=&\frac 14\varepsilon _{\mu \nu \rho }F^{\nu \rho
}(x)\int d^3y\left| x-y\right| F^2(y)\;,  \label{nl-coeff} \\
\vartheta _\mu ^2 &=&-\frac 18\left( \int d^3y\left| x-y\right|
F^2(y)\right) D_\sigma ^x\int d^3zF_\mu ^{\;\sigma }(x)\left| x-z\right|
F^2(z)\;,  \nonumber
\end{eqnarray}
$D_\sigma ^x\;$being the covariant derivative acting on the point $x.\;$Thus

\begin{equation}
\mathcal{S}_{CS}(A)+\mathcal{S}_{YM}^{nloc}(A)=\;\mathcal{S}_{CS}(\widehat{A}%
)+O(1/m^3)\;,  \label{nl-red}
\end{equation}
with 
\begin{equation}
\widehat{A}_\mu =A_\mu +\frac 1m\vartheta _\mu ^1+\frac 1{m^2}\vartheta _\mu
^2+O(1/m^3)\;.  \label{nl-conn}
\end{equation}
As anticipated, the coefficients $\vartheta _\mu ^1,\vartheta _\mu ^2$ in eq.%
$\left( \text{\ref{nl-coeff}}\right) $, although nonlocal, transform
covariantly\footnote{%
Perhaps the covariant character of the coefficients $\vartheta _\mu $ in the
nonlocal case could be understood by noticing that the Lemma of the Section
3, being of a purely geometrical nature, could be in principle applied to
nonlocal actions built up with $F$ and its covariant derivatives.}, so that
the redefined field $\widehat{A}_\mu $ is \textit{still a connection}. It is
worth emphasizing that the form of the space-time function between $F^2(x)$
and $F^2(y)$ in the eq.$\left( \text{\ref{nl-ym}}\right) $ is in fact
completely irrelevant. More sophisticated examples of nonlocal $F$-dependent
actions can be worked out, leading to similar results. We see therefore that
a large class of nonlocal terms can be reabsorbed in the pure Chern-Simons.
In our opinion this observation is a signal of the fact that the
aforementioned rigidity of the topological Chern-Simons may persist at the
quantum level. Although we have a solid understanding of the BRST cohomology
in the space of the local functionals \cite{bdk,dv,bbh}, the situation is
completely different in the nonlocal case. Up to our knowledge there is no
proof of the fact that the nonlocal terms are built essentially with the
field strength and its derivatives. Notice that we are not demanding here
the complete characterization of the nonlocal terms, including the knowledge
of the space-time dependence of the $1PI$ $n$-point Green function $\Gamma
^n(x_{1,....,}x_n)$. A weaker characterization guarantying the simple
presence of $F$ would be almost sufficient in order to establish the quantum
equivalence between $\Gamma (A)$ and $\mathcal{S}_{CS}(A).$ It would be a
very nice event if in the case of the flat space-time $\mathcal{R}^3$ the $%
1PI$ effective action could be resetted to a pure Chern-Simons, up to
nonlinear field redefinitions.

Let us conclude this section by drawing a possible path in favour of this
hypothesis. The forthcoming considerations heavily rely on a rather
appealing suggestion of \cite{pr} (see in particular Sect.IV) and on the
explicit one and two loop computations on topological massive Yang-Mills and
on pure Chern-Simons done till now \cite{tmym,pr,gmrr,as,gl,en,ag}.

After having reabsorbed all the nonlocal terms that we can, we should be
able to write the complete $1PI$ effective action $\Gamma (A)$ in the
following form

\begin{equation}
\Gamma (A)=\zeta \mathcal{S}_{CS}(\widehat{A})\;+\Xi \;,  \label{dream}
\end{equation}
with 
\begin{equation}
\widehat{A}=A+\delta A^{loc}+\delta A^{nloc}\;.  \label{dream1}
\end{equation}
The coefficient\footnote{%
There has been a long discussion in the last years about a possible
universal meaning of $\zeta $. This question being not addressed here, we
remind the original literature \cite{g1,as,ag}.} $\zeta $ in eq.$\left( 
\text{\ref{dream}}\right) $ is a power series in $\hbar $ accounting for
possible finite corrections to the Chern-Simons term itself and can be
reabsorbed by a further \textit{finite} \textit{multiplicative }redefinition
of the gauge field and of the coupling constant $g.\;$

\noindent The extra term $\Xi $ in eq.$\left( \text{\ref{dream}}\right) $
represents all the gauge invariant nonlocal terms which cannot be reabsorbed
through nonlinear field redefinitions. However $\Xi $ should be very
constrained. It should not contain $m$ since, due to the eqs.$\left( \text{%
\ref{q-e-m}}\right) $\footnote{%
It is useful to recall here that the action $\Sigma $ in eq.$\left( \ref
{ca-l}\right) $ differs from the pure massive topological Yang-Mills by the
infinite set of gauge parameters $\alpha _k^j$.}, $m$-dependent terms are
expected to be related to nonlinear redefinitions. Then $\Xi $ should
survive then the infinite mass limit $m\rightarrow \infty .$ Pure
Chern-Simons considerations should thus apply. Therefore, according to \cite
{pr} (see Sect.IV), $\Xi $ should be vanishing, due to a parity argument 
\cite{pr} and to the general covariance of the Chern-Simons in the Landau
gauge \cite{en,ag,al,noi,lu}. In summary, the $1PI$ effective action of
topological Yang-Mills in flat space-time $\mathcal{R}^3$ should be resummed
to pure Chern-Simons, up to nonlinear field redefinitions. This behaviour
may be completely different for a generic curved three manifold, as other
kinds of topological invariants like the Ray-Singer torsion \cite{rs} are
expected to appear.

It is rather important to underline here that a nonlocal field redefinition
has been in fact already used by \cite{gl} in order to reset the one loop
effective action of Chern-Simons in the light cone gauge to a pure
Chern-Simons action.

Other kinds of three dimensional effective actions, as for instance the
fermionic determinant of a two component massive spinor could be
contemplated. Indeed, the infinite mass limit of the abelian fermionic
determinant in flat space-time is nothing but pure Chern-Simons \cite{det}$.$
Moreover, the perturbative expression obtained so far for the abelian
determinant can be easily reabsorbed through nonlinear redefinitions \cite
{prep} into a pure Chern-Simons, providing thus a further evidence

\subsection{Final remarks}

We emphasize once again that the interpretation of the Chern-Simons as a
gauge invariant functional which is able to reproduce any given \textit{%
local }Yang-Mills type action is rather attractive.

Whether this pure geometrical set up can be extended at the level of the
full $1PI$ effective quantum action is still an open question. In any case a
great part of $\Gamma (A)\;$can be certainly resetted to a pure Chern-Simons
term up to nonlinear field redefinitions.

It is worth underlining that the covariant character of the coefficients $%
\vartheta _\mu $ (in both local and nonlocal case) which allows to interpret
the redefined field $\widehat{A}_\mu $ as a gauge connection could shed some
light on the role of the nonlinear field redefinitions in quantum field
theory. This point could be of a certain relevance within the new recent
perspectives on the renormalization of gauge theories \cite{gw}.

The present results naturally remind us the relationship with general
relativity. After all, being $\widehat{A}_\mu $ a connection, the resulting
action $\mathcal{S}_{CS}(\widehat{A})$ is gauge invariant and looks as good
as the initial one $\mathcal{S}_{CS}(A)$. This suggests that, as far as the
gauge invariance is taken as the guide principle in order to select
appropriate actions, we should still have the freedom of choosing the
connection.

The inclusion of matter fields is under investigation.

We hope, finally, that this work will be of some help in order to improve
our present understanding of three dimensional gauge theories and of the
role of the topological actions.

\vspace{5mm}

{\Large \textbf{Acknowledgements}}

We wish to express our gratitude to R. Jackiw for a comment on our previous
letter. We are indebted to D.G. Barci, F. Fucito, J.A. Helay\"{e}l-Neto, M.
Henneaux, M. Martellini, O. Piguet and G. Oliveira Pires for fruitful
discussions.

The Conselho\ Nacional de Pesquisa e Desenvolvimento, CNP$q$ Brazil, the
Faperj, Funda\c {c}\~{a}o de Amparo \`{a} Pesquisa do Estado do Rio de
Janeiro and the SR2-UERJ are gratefully acknowledged for financial support.

\vspace{5mm}

\appendix

\section{Appendix}

\subsection{Extended BRST technique}

The extended BRST technique \cite{co} is a very powerful tool which allows
to control the dependence of the theory at the quantum level from parameters
associated to exact BRST terms. Let us present here how this technique works
in the case of the parameter $m$ of the action $\Sigma $ in eq.$\left( \text{%
\ref{ca-l}}\right) $

\begin{equation}
\Sigma =\mathcal{S}+tr\int d^3x\left( b\partial A+\partial ^\mu \overline{c}%
D_\mu c+A_\mu ^{*}D^\mu c-gc^{*}c^2\right) \;,  \label{r-ca-l}
\end{equation}
with

\begin{equation}
\mathcal{S}(A)=\mathcal{S}_{CS}(A)+\frac 1{4m}tr\int d^3xF_{\mu \nu }F^{\mu
\nu }+\sum_{j=2}^\infty \frac 1{m^j}\left( \sum_{k=1}^{d_k}\alpha _k^j%
\mathcal{S}_j^k(A)\right) \;.  \label{r-m-g}
\end{equation}
The action $\Sigma $ obeys the classical Slavnov-Taylor identity

\begin{equation}
tr\int d^3x\left( \frac{\delta \Sigma }{\delta A_\mu }\frac{\delta \Sigma }{%
\delta A^{*\mu }}+\frac{\delta \Sigma }{\delta c}\frac{\delta \Sigma }{%
\delta c^{*}}+b\frac{\delta \Sigma }{\delta \overline{c}}\right) =0\;,
\label{s-t}
\end{equation}
from which it follows that the so called linearized operator $\mathcal{B}%
_\Sigma $ defined as

\begin{equation}
\mathcal{B}_\Sigma =tr\int d^3x\left( \frac{\delta \Sigma }{\delta A_\mu }%
\frac \delta {\delta A^{*\mu }}+\frac{\delta \Sigma }{\delta A^{*\mu }}\frac
\delta {\delta A_\mu }+\frac{\delta \Sigma }{\delta c}\frac \delta {\delta
c^{*}}+\frac{\delta \Sigma }{\delta c^{*}}\frac \delta {\delta c}+b\frac
\delta {\delta \overline{c}}\right) \;,  \label{l-s-t}
\end{equation}
is nilpotent

\begin{equation}
\mathcal{B}_\Sigma \mathcal{B}_\Sigma =0\;.  \label{n}
\end{equation}
As it is well known \cite{book}, this operator identifies the full BRST
differential acting on the fields and antifields.

\noindent Owing to the results of Sects.3,4, we have

\begin{equation}
\frac{\partial \Sigma }{\partial m}=\mathcal{B}_\Sigma \Lambda \;,
\label{m-triv}
\end{equation}
$\Lambda $ being an integrated local formal power series in the fields and
antifields of ghost number -1. According to \cite{co}, we introduce the term 
$\Lambda $ in the classical action $\Sigma $ by means of a constant
parameter $\xi $ of ghost number 1, namely

\begin{equation}
\widetilde{\Sigma }=\Sigma +\xi \Lambda \;.  \label{act-h}
\end{equation}
Let us now compute the quantity

\begin{equation}
tr\int d^3x\left( \frac{\delta \widetilde{\Sigma }}{\delta A_\mu }\frac{%
\delta \widetilde{\Sigma }}{\delta A^{*\mu }}+\frac{\delta \widetilde{\Sigma 
}}{\delta c}\frac{\delta \widetilde{\Sigma }}{\delta c^{*}}+b\frac{\delta 
\widetilde{\Sigma }}{\delta \overline{c}}\right) \;.  \label{q}
\end{equation}
The expression $\left( \text{\ref{q}}\right) $ is expected to be
nonvanishing, due to the use of the modified action $\widetilde{\Sigma }$.
However, due to the fact that $\xi \xi =0,$ we easily get

\begin{equation}
tr\int d^3x\left( \frac{\delta \widetilde{\Sigma }}{\delta A_\mu }\frac{%
\delta \widetilde{\Sigma }}{\delta A^{*\mu }}+\frac{\delta \widetilde{\Sigma 
}}{\delta c}\frac{\delta \widetilde{\Sigma }}{\delta c^{*}}+b\frac{\delta 
\widetilde{\Sigma }}{\delta \overline{c}}\right) =-\xi \mathcal{B}_\Sigma
\Lambda \;.  \label{b-s-t}
\end{equation}
Therefore from 
\begin{equation}
\xi \mathcal{B}_\Sigma \Lambda =\xi \frac{\partial \Sigma }{\partial m}=\xi
\left( \frac{\partial \widetilde{\Sigma }}{\partial m}-\xi \frac{\partial
\Lambda }{\partial m}\right) =\xi \frac{\partial \widetilde{\Sigma }}{%
\partial m}\;,  \label{nice}
\end{equation}
it follows that the action $\widetilde{\Sigma }$ satisfies the modified
Slavnov-Taylor identity

\begin{equation}
\int d^3xtr\left( \frac{\delta \widetilde{\Sigma }}{\delta A_\mu }\frac{%
\delta \widetilde{\Sigma }}{\delta A^{*\mu }}+\frac{\delta \widetilde{\Sigma 
}}{\delta c}\frac{\delta \widetilde{\Sigma }}{\delta c^{*}}+b\frac{\delta 
\widetilde{\Sigma }}{\delta \overline{c}}\right) +\xi \frac{\partial 
\widetilde{\Sigma }}{\partial m}=0\;.  \label{m-s-t}
\end{equation}
This expression is easily recognized to be of the type of an extended
Slavnov-Taylor identity \cite{co}. In fact, from a cohomological point of
view the parameters $\xi ,m$ form a doublet, \textit{i.e.}

\begin{equation}
\mathcal{B}_{\widetilde{\Sigma }}m=\xi \;,\;\;\;\;\;\;\;\mathcal{B}_{%
\widetilde{\Sigma }}\xi =0\;.  \label{doub}
\end{equation}
The absence of gauge anomalies in three dimensions (see Subsect.2.2)
guaranties thus that the identity $\left( \text{\ref{m-s-t}}\right) $ holds
at the quantum level

\begin{equation}
\int d^3xtr\left( \frac{\delta \widetilde{\Gamma }}{\delta A_\mu }\frac{%
\delta \widetilde{\Gamma }}{\delta A^{*\mu }}+\frac{\delta \widetilde{\Gamma 
}}{\delta c}\frac{\delta \widetilde{\Gamma }}{\delta c^{*}}+b\frac{\delta 
\widetilde{\Gamma }}{\delta \overline{c}}\right) \;+\xi \frac{\partial 
\widetilde{\Gamma }}{\partial m}=0\;.  \label{q-m-s-t}
\end{equation}
Acting now on the expression $\left( \text{\ref{q-m-s-t}}\right) \;$with the
test operator $\partial /\partial \xi $ and setting $\xi $ to zero we
immediately get

\begin{eqnarray}
\frac{\partial \Gamma }{\partial m} &=&\mathcal{B}_\Gamma \left[ \Lambda
\cdot \Gamma \right] \;,  \label{res} \\
\Gamma &=&\left. \widetilde{\Gamma }\right| _{\xi =0}\;,  \nonumber
\end{eqnarray}
thereby proving the statement $\left( \text{\ref{q-e-m}}\right) .\;$ The
same procedure applies to the parameters $\alpha _k^j$ as well as to the
energy-momentum tensor. In the latter case the extended BRST technique has
to be done twice. First one considers the integrated insertion $\int
d^3x\left[ T_{\;\mu }^\mu \cdot \Gamma \right] $, for which one gets

\begin{equation}
\int d^3x\left[ T_{\;\mu }^\mu \cdot \Gamma \right] =BRST-\mathrm{%
variation\;.}  \label{first}
\end{equation}
A further application of the extended BRST technique with local space-time
dependent parameters allows to obtain the final result $\left( \text{\ref
{q-e-m}}\right) .$


\vspace{1.0in}

\begin{thebibliography}{99}
\bibitem{lett}  V.E.R. Lemes, C. Linhares de Jesus, C.A.G. Sasaki, S.P.
Sorella, L.C.Q. Vilar and O.S. Ventura, \textit{A simple remark on three
dimensional gauge theories, }\textbf{CBPF-NF-050/97, hep-th/9708098, }to
appear in \textbf{Phys. Lett. B};

\bibitem{tmym}  R. Jackiw and S. Templeton, \textbf{Phys. Rev. D23 (1981)
2291;}\\S. Deser, R. Jackiw and S. Templeton, \textbf{Ann. Phys. (N.Y.) 140
(1982) 372;}\\S. Deser, R. Jackiw and S. Templeton,\ \textbf{Phys. Rev.
Lett, 48 (1982) 975;}

\bibitem{bdk}  F. Brandt, N. Dragon and M. Kreuzer, \textbf{Phys. Lett. B231
(1989) 263;}\\F. Brandt, N. Dragon and M. Kreuzer, \textbf{Nucl. Phys. B332
(1990) 224;}\\F. Brandt, N. Dragon and M. Kreuzer, \textbf{Nucl. Phys. B332
(1990) 250;}

\bibitem{dv}  M. Dubois-Violette, M. Henneaux, M. Talon and C.M. Viallet, 
\textbf{Phys. Lett. B289 (1992) 361;}

\bibitem{bbh}  G. Barnich, F. Brandt and M. Henneaux, \textbf{Comm. Math.
Phys. 174 (1995) 57;}\\G. Barnich, F. Brandt and M. Henneaux, \textbf{Comm.
Math. Phys. 174 (1995) 93;}

\bibitem{ht}  M. Henneaux and C. Teitelbom, \textit{Quantization of Gauge
System, }\textbf{Priceton University Press 1992;}

\bibitem{book}  O. Piguet and S.P. Sorella, \textit{Algebraic
Renormalization, }\textbf{Springer-Verlag, Berlin, 1995;}

\bibitem{dix}  J. Dixon, \textit{Cohomology and renormalization of gauge
theories, }I,II,III, \textbf{unpublished;}\\J. Dixon, \textbf{Comm. Math.
Phys. 139 (1991) 495;}

\bibitem{bh}  G. Barnich and M. Henneaux, \textbf{Phys. Lett. B 311 (1993)
123;}

\bibitem{bbrt}  D. Birmingham, M. Blau, M. Rakowski and G. Thompson, \textbf{%
Phys. Rep. 209 (1991) 129};

\bibitem{g1}  G. Giavarini, C.P. Martin, F. Ruiz Ruiz, \textbf{Phys. Lett. B
314 (1993) 328;}\\G. Giavarini, C.P. Martin, F. Ruiz Ruiz, \textbf{Phys.
Lett. B 332 (1994) 345;}

\bibitem{as}  M. Asorey, F. Falceto, J.L. L\'{o}pez and G. Luz\'{o}n, 
\textbf{Nucl. Phys. B429 (1994) 344;}\\M. Asorey, F. Falceto, J.L. L\'{o}pez
and G. Luz\'{o}n,\textbf{\ Phys. Rev. D49 (1994) 5377;}

\bibitem{pr}  R.D. Pisarski and S. Rao, \textbf{Phys. Rev. D32 (1985) 2081;}

\bibitem{gmrr}  G. Giavarini, C.P. Martin, F. Ruiz Ruiz, \textbf{Nucl. Phys.
B 381 (1992) 222;}

\bibitem{kp}  G.A.N. Kapustin and P.I. Pronin, \textbf{Phys. Lett. B 318
(1993) 465;}\\G.A.N. Kapustin and P.I. Pronin, \textbf{Mod. Phys. Lett. A9
(1994) 1925;}

\bibitem{gw}  J. Gomis and S. Weinberg, \textbf{Nucl. Phys. B 469 (1996) 473;%
}

\bibitem{sigma}  C. Becchi, A. Blasi, G. Bonneau, R. Collina and F. Delduc, 
\textbf{Comm. Math. Phys. 120 (1988) 121;}

\bibitem{n1}  O. Piguet and K. Sibold, \textit{Renormalized Supersymmetry , }%
series \textit{Progress in Physics , }\textbf{vol. 12, Birkh\"{a}user Boston
Inc.,}\textit{\ }\textbf{1986;}

\bibitem{co}  O. Piguet and K. Sibold, \textbf{Nucl. Phys. B253 (1985) 517;}

\bibitem{rec}  O.M. Del Cima, D.H.T. Franco, J.A. Helay\"{e}l-Neto and O.
Piguet,\textit{\ On the Non-Renormalization Properties of Gauge theories
with Chern-Simons Terms, }\textbf{CBPF-NF-052/97, UFES-DF-OP97/2,
hep-th/9711191;}

\bibitem{en}  E. Guadagnini, M. Martellini and M. Mintchev, \textbf{Phys.
Lett. B224 (1989) 489;}\\E. Guadagnini, M. Martellini and M. Mintchev, 
\textbf{Phys. Lett. B227 (1989) 111}\\P. Cotta Ramusino, E. Guadagnini, M.
Martellini and M. Mintchev, \textbf{Nucl. Phys. B330 (1990) 557;}\\E.
Guadagnini, M. Martellini and M. Mintchev, \textbf{Nucl. Phys. B330 (1990)
575;}

\bibitem{ag}  L. Alvarez-Gaum\'{e}, J.M.F. Labastida and A.V. Ramallo, 
\textbf{Nucl. Phys. B334 (1990) 103;}

\bibitem{al}  A. Blasi and R. Collina, \textbf{Nucl. Phys. B345 (1990) 472;}

\bibitem{noi}  F. Delduc, C. Lucchesi, O. Piguet and S. P. Sorella, \textbf{%
Nucl. Phys. B346 (1990) 513;}

\bibitem{gl}  G. Leibbrandt and C.P. Martin, \textbf{Nucl. Phys. B377 (1992)
593;}\\G. Leibbrandt and C.P. Martin, \textbf{Nucl. Phys. B416 (1994) 351;}

\bibitem{rs}  A. Schwarz, \textbf{Lett. Math. Phys. 2 (1978) 247;}

\bibitem{det}  A. Niemi and G.W. Semenoff, \textbf{Phys. Rev. Lett, 23
(1983) 2077};\\A. N. Redlich, \textbf{Phys. Rev. D29 (1984) 2366;\\}A. Coste
and M. L\"{u}scher; \textbf{Nucl. Phys. B323 (1989) 631;}

\bibitem{lu}  C. Lucchesi and O. Piguet, \textbf{Nucl. Phys. B381 (1992) 281;%
}

\bibitem{prep}  D. G. Barci et. al., \textit{forthcoming paper}.
\end{thebibliography}
\end{document}